\documentclass[12pt, a4paper]{article}
\usepackage[utf8]{inputenc}
\usepackage{hyperref}
\usepackage[centertags]{amsmath}
\usepackage{bm}       
\usepackage{graphicx}
\usepackage{xcolor}
\usepackage{pdflscape}
\let\latexput\put
\usepackage{pictexwd}
\usepackage{array}
\usepackage{color}
\usepackage{multirow}
\usepackage{tablefootnote}
\usepackage{longtable}
\usepackage{verbatim}
\usepackage{authblk} 

\newcolumntype{L}[1]{>{\raggedright\let\newline\\\arraybackslash\hspace{0pt}}p{#1}}
\newcolumntype{C}[1]{>{\centering\let\newline\\\arraybackslash\hspace{0pt}}p{#1}}
\newcolumntype{R}[1]{>{\raggedleft\let\newline\\\arraybackslash\hspace{0pt}}p{#1}}

\newcommand{\ccod}[1]{CCOD#1}
\newcommand{\ssd}[1]{SSD#1}

\let\put\latexput
\title{Clinical trials with interim analyses: Standardizing Terminology to increase clarity}

\author[1]{Elina Asikanius\footnote{Asikanius and Hofner share first authorship. Kunz and Rufibach share last authorship. }}
\author[2, 3]{Benjamin Hofner$^{*}$}
\author[4]{Lisa V Hampson}
\author[5]{Gernot Wassmer}
\author[6]{Christopher Jennison}
\author[7]{Tobias Mielke}
\author[8]{Cornelia Ursula Kunz$^{*}$}
\author[9]{Kaspar Rufibach$^{*}$}

\affil[1]{Finnish Medicines Agency, Turku, Finland}
\affil[2]{Section Data Science and Methods, Paul-Ehrlich-Institut, Langen, Germany}
\affil[3]{Department of Medical Informatics, Biometry and Epidemiology, Friedrich-Alexander-Universität Erlangen-Nürnberg, Erlangen, Germany}
\affil[4]{Advanced Methodology \& Data Science, Novartis Pharma AG, Basel, Switzerland}
\affil[5]{RPACT GbR, Am Rodenkathen 11, Sereetz, 23611, Germany}
\affil[6]{Department of Mathematical Sciences, University of Bath, Bath, UK}
\affil[7]{Statistics and Decision Sciences, Janssen-Cilag GmbH, Neuss, Germany}
\affil[8]{Methodology Group, Global Biostatistics and Data Sciences, Boehringer-Ingelheim, Biberach, Germany}
\affil[9]{Methods, Collaboration, and Outreach Group, Product Development Data Sciences (MCO), Hoffmann-La Roche Ltd, Basel, Switzerland}

\date{}

\begin{document}

\maketitle

\begin{abstract}
Interim analyses for group-sequential decision making are prevalent in clinical trials. Methodology is well established and has been routinely implemented over the last decades. Still, confusions and uncertainties on aspects of how to operationalize and interpret interim analyses exist for many stakeholders. In this paper, a team of statisticians from the pharmaceutical industry, academia, and regulatory agencies provide a multi-stakeholder perspective on the key concepts behind interim analyses, with the aim to introduce standard terminology to mitigate misunderstandings and facilitate clearer discussions. 
\end{abstract}

\subsection*{Disclaimer}
The views expressed in this article are the personal views of the authors and may not be understood or quoted as being made on behalf of or reflecting the position of the regulatory agencies or organizations with which the authors are employed/affiliated.

\section{Introduction}
\label{sec:intro}

During the past decades, interim analyses implemented through a group-sequential or adaptive design have become common \cite{Bothwelle018320}, especially for trials with time-to-event endpoints. The statistical theory is well established and straightforward to implement at the design stage. Terminology for such designs is, however, not standard, and to some extent, also misleading. Here, as a team of statisticians from pharmaceutical industry, regulatory agencies, and academia, we propose terminology for clinical trials with interim analyses and confirmatory intent, i.e. hypothesis testing. Interim analyses are defined as pre-specified analyses that are performed during a clinical trial to determine whether criteria are met to implement pre-planned adjustments to the trial design \cite{fda_19_adaptive,ema_adaptive,iche9}. Often, such criteria refer to potentially reaching an early conclusion on a scientific hypothesis. Different roles for interim analyses can be considered: (1) Interim analyses for efficacy, (2) interim analyses for futility, (3) interim analyses for safety and (4) interim analyses in adaptive designs where trial design features can be adapted, e.g., an arm is dropped, the sample size of the trial is modified, or one of several pre-specified populations is selected for enrichment. At a given interim analysis one or more of these assessments can be made.

We restrict our attention to group-sequential designs, by which we mean the subclass of adaptive designs where either futility or efficacy can be claimed at an interim analysis, but no other modifications of the trial are foreseen by design \cite{wassmer_16}. The intent of an efficacy interim analysis is to reject the null hypothesis that the treatment has no benefit over the control treatment and claim efficacy; the intent of a futility analysis is to stop the trial if early data suggests that the trial is highly unlikely to reject the null hypothesis at later analyses. Naturally, both of these actions (stopping for futility or efficacy) can be considered at a single interim analysis time point.

An interim analysis has implications for the conduct, reporting, and interpretation of a clinical trial. In our experience working as statisticians at regulatory agencies, in pharmaceutical companies, or in academia, we have observed that the design and statistical analysis plans of trials with interim analyses are at times not appropriately outlined and therefore the trials are inadequately reported. Guidance on how to write statistical analysis plans exists, see e.g. Gamble et al.\cite{gamble_17}. However, these guidances are not always followed and also do not require all the aspects that we propose to introduce for clearer communication. All this makes trial designs and their results prone to misinterpretation within the trial team and by relevant external stakeholders such as regulatory agencies, HTA (Health Technology Assessment) bodies, health care providers, investigators, patients and potentially even the wider public: an example of such misconceptions and misinterpretations based on the first four COVID-19 vaccines licensed in the EU is described by Hofner et al.\cite{Hofner2023}.
This paper targets all stakeholders of clinical trials with interim analyses, i.e., statisticians and clinicians in the pharmaceutical industry, regulatory agencies including HTA bodies, and academia. It is essential that all parties involved in planning, conduct, and interpretation of a clinical trial fully understand all the aspects related to the interim analysis strategy and that these are communicated in a transparent and precise manner to internal and external stakeholders. Currently, our impression is that there is a lack of standardized terminology, including across pertinent (Health Authority) guidelines, which complicates the interpretation and clear communication of clinical trial results. Furthermore, the possible outcomes of interim analyses and their impact on trial conduct are often not clearly outlined. Through this paper, we recommend standard terminology and clarify aspects related to trial conduct and potential trial outcomes. This paper will facilitate: 1) informed decision making at the design stage; 2) trial conduct so that all involved parties have a full understanding of the intended scope of analyses; and 3) alignment and therefore a common interpretation of results across internal and external stakeholders.
This work complements Ciolino et al (2023)\cite{ciolino_23} who provide a general overview and guidance on interim analyses for a nonstatistical audience. Based on our experience as various stakeholders involved in clinical trials at large, we focus on aspects of aligning terminology and clarifying communication of designs and results of clinical trials with interim analyses. As such, we target drug developers in general, including statisticians working on clinical trials.
Section~\ref{sec:example} introduces general concepts through a hypothetical trial that we then use in Section~\ref{sec:general} to illustrate the core of the paper, our proposed general terminology, which we present in Tables~\ref{tab: glossary} and \ref{tab: glossary2}. Additional aspects are outlined in Section~\ref{sec:add}. We conclude with a discussion in Section\ref{sec:example_protocol}. We provide an example of text describing a trial design that could appear in a protocol or statistical analysis plan in an appendix to the paper. Also an example to illustrate reporting of such a trial is given in the appendix.

\section{Motivating Example -- a hypothetical trial}
\label{sec:example}

We consider a hypothetical randomized controlled trial (RCT) with two treatment arms, experimental (E) vs. control (C). Let the primary endpoint be a time-to-event outcome, such as progression-free survival (PFS) or overall survival (OS) in oncology, or time to first relapse in multiple sclerosis. We shall comment on more general endpoint definitions, multi-armed situations, and subgroup analyses in Section~\ref{sec:discussion}.

Using a one-sided significance level of 0.025, suppose we want to have 80\% power to reject the null hypothesis of no benefit of treatment when the true hazard ratio is 0.75. Without an interim analysis we would need to target 380 events to have a trial with these specifications. A minimal detectable difference of 0.8177 would need to be observed for a statistically significant result at the primary analysis. We plan to recruit 1200 patients over approximately 12 months. 

The term {\it minimal detectable difference} refers to the minimal observed treatment effect which can still become statistically significant with the given sample size, or in other words the critical value of the hypothesis test on the scale of interest, here the hazard ratio. It may refer to an actual difference, e.g. between means, or to a relative risk measure such as the hazard ratio. To avoid statistically significant results that are not clinically relevant it is of importance to discuss with trial teams whether the minimal detectable difference is clinically relevant (0.8177 in our example), not necessarily whether the effect we power at (0.75 in our example) is clinically relevant. This has also been emphasized precisely in the context of a time-to-event endpoint in drug development by Carroll\cite{carroll_09}, where also formulas for minimal detectable difference computation for a time-to-event endpoint are provided. For the minimal detectable difference in the context of mean differences see, for example, Brock et al.\cite{brock_15}.

To reduce the \textit{expected} number of events (and hence to reduce the expected time needed to have the results available) in the case that the experimental treatment does indeed work, we plan an interim analysis for efficacy at $2/3$ of the targeted number of events events. We use a Lan-DeMets $\alpha$-spending approximation~\cite{lan_83} to O'Brien-Fleming~\cite{obrien_79} boundaries to control the overall type I error. To maintain the overall power, the maximum number of events is increased to 385. Consequently, with this interim analysis planned at an information fraction of $2/3$, it should be conducted after 257 of 385 events.

Similarly, to reduce the expected number of events if the intervention is not efficacious, we add non-binding interim analyses for futility (see discussion in Section~\ref{sec:stopping}), the first after an information fraction of 1/3 (IA1, after 129 events) and the second after an information fraction of 2/3 (IA2, after 257 events), i.e., at the same time as the efficacy interim analysis. An independent data monitoring committee (iDMC, see below) is instructed to recommend stopping the trial for futility based on the observed hazard ratio if:
\begin{itemize}
    \item at IA1 the observed hazard ratio is $\geq 1$, or 
    \item at IA2 the observed hazard ratio is $\geq 0.9$.
\end{itemize}
Assuming these decision criteria are followed, adding the futility analyses reduces the overall power of the trial from 80 to 78\%. This is a direct consequence of the fact that even if the assumption on the alternative hypothesis is true, there are cases -- albeit rather unlikely -- where at the interim analyses the futility criterion would be met. It would be straightforward to compensate for this power loss by increasing the targeted number of events. For simplicity, and as the loss of power is minimal, the primary analysis remains planned at 385 events. Section~\ref{sec:example_protocol} provides some more details and illustrates how we suggest to specify these interim analyses in a trial protocol or statistical analysis plan. 

To ensure smooth conduct of the trial, the interim analyses should be carefully pre-planned and implications of interim decisions should be understood by all parties involved. In order to maintain trial integrity, interim analyses in registrational trials are typically conducted by an iDMC~\cite{iche9}. The iDMC will provide recommendations to the sponsor of the trial. Guidelines for those recommendations are typically described clearly in an iDMC-charter. If need be, the iDMC may deviate from the proposed recommendation guidelines, to recommend alternative actions (i.e. recommendation to add another interim analysis). Any recommendations from the iDMC are reduced to a minimum informative level (continue the trial unchanged, stop the trial for success) and are directed towards a sponsor committee, which itself is limited in size for integrity reasons. The sponsor may then follow the recommendations - or not. Further guidance on the installation and role of iDMCs are available from the FDA~\cite{fda_idmc} and the EMA~\cite{ema_DMC_GL, ema_DMC_QA}, see also Sch\"offski\cite{schoeffski_21}. Fleming et al.\cite{fleming_17} provide additional guidance on best practices for iDMCs based on extensive experience.

General terminology for group-sequential designs, which will be helpful to align at the design stage of the trial, is provided in Tables~\ref{tab: glossary} and \ref{tab: glossary2}. Many of the terms in these tables (e.g., first-patient-in, clinical cutoff date \ccod, and snapshot data \ssd) are also applicable to trials with other types of endpoints and without an interim analysis. Note that \ccod{} and \ssd{} are {\it analysis-specific}. For each analysis, a corresponding \ccod{} and \ssd{} is defined and the corresponding dataset is only generated at the \ssd.

Notably, the simple term {\it sample size} may be confusing in communications across stakeholders. The {\it target sample size} in a group-sequential design with a binary or continuous endpoint is the maximum number of {\it patients}. In trials with time-to-event endpoints, it is recommended to specifically use the terms {\it number of events} and {\it number of patients} and in fact an interim analysis does not usually affect the number of patients in the trial. The term {\it sample size} is a frequent source for confusion between the former and the latter.

\begin{landscape}
\begin{table}[htbp]
\footnotesize
\centering
\begin{tabular}{|L{3cm}|L{9cm}|L{9cm}|} \hline
{\bf Term}                        & {\bf Definition} & {\bf Illustrated using hypothetical trial} \\ \hline
Group-sequential design (GSD)
& A group-sequential design is a special adaptive design which allows for one or more interim analyses where the trial can potentially be stopped for efficacy, with rejection of the null hypothesis, or for futility, when the null hypothesis is unlikely to be rejected if the trial continues.  
& See Table~\ref{tab:trial1} for the design details of our hypothetical trial. \\ 
\hline
First patient in
&  The date at which the first patient enters a clinical trial. In a randomized controlled trial typically the date when the first patient is randomized. 
& We assume our first patient was randomized on 23rd April 2020. \\ 
\hline 
Cutoff decision date 
& For a trial with a time-to-event endpoint the date at which the trial team decides when the next clinical cutoff date will be. The decision is based on a projection on when the targeted number of events will be reached based on the history of (timing of) events observed so far.
& See Section~\ref{subsec:running_interim}. \\ 
\hline
Clinical cutoff date (\ccod)
& Data recorded on or before this date are included in the analysis. 
& The \ccod{} for IA1 has been {\it predicted} to be 19.7 months after first patient in, so 14th December 2021.  For more on predicting such dates, see Section~\ref{subsec:running_interim}. The actual \ccod{} may be different and will depend on the accrual of events (time-to-event endpoints). Any event up to the \ccod{} will be included in the analysis.\\ 
\hline
Snapshot date (\ssd)
& Date at which the datasets that will be used for the analysis are ready for extraction. The \ssd{} differs from the \ccod{} as trial sites need to enter all relevant data which has accrued until the \ccod{} and this data usually needs to be cleaned and and/or adjudicated by the sponsor.
& Assume it takes six weeks for investigating sites to enter their data and for the sponsor to clean it. Then the \ssd{} would be six weeks after the \ccod, i.e., on 28th January 2022. On that date all data which accrued on or before the \ccod{} would be extracted from the database. Events which occur after \ccod{} but prior to \ssd{} will not be included in this analysis.\\ \hline
\end{tabular}
\caption{Glossary of general terms used in this paper, Part 1. }
\label{tab: glossary}
\end{table}
\end{landscape}

\begin{landscape}
\begin{table}[htbp]
\footnotesize
\centering
\begin{tabular}{|L{3cm}|L{9cm}|L{9cm}|} \hline
{\bf Term}                        & {\bf Definition} & {\bf Illustrated using hypothetical trial} \\ \hline
Target information
& GSDs are designed assuming a maximum number of patients (binary and continuous endpoint) or a maximum number of events (time-to-event endpoint). This maximum number is the target information (also referred to as {\it target number of patients} or {\it target number of events}).     
& Our hypothetical trial has a target number of events of 385, see Table~\ref{tab:trial1}. \\
\hline 
Information fraction
& The proportion of patients or events available at an interim analysis relative to the target number of patients or events. 
& See Column {\it Information Fraction} in Table~\ref{tab:trial1}. \\ \hline 
Independent data monitoring committee 
& An iDMC is set up to ensure the integrity of the clinical trial. The iDMC has access to potentially unblinded data and gives recommendations based on pre-defined rules. 
& \\
\hline
\end{tabular}
\caption{Glossary of general terms used in this paper, Part 2. }
\label{tab: glossary2}
\end{table}
\end{landscape}

\section{Terminology for Different Types of Analyses}
\label{sec:general}

Moving through design, implementation and finally reporting of group-sequential designs, confusion can be generated through inappropriate use of seemingly ``simple'' standard terms, such as {\it interim analysis}, {\it primary analysis} or {\it final analysis}. In this section, we discuss some potential sources for confusion and propose alignment of terminology.
Tables~\ref{tab: glossary_analyses1} and \ref{tab: glossary_analyses2} define terms that relate to different analyses in group-sequential designs. Similar terms were previously defined by Hofner et al.~\cite{Hofner2023} with the aim of demonstrating that standardized terminology would have enabled better informed discussions about the COVID-19 vaccine trials amongst all stakeholders. 
The first important distinction to draw is between the purpose of an analysis as envisaged at the {\it design phase}, and the actual purpose it has at the {\it reporting stage} of a trial. Furthermore, some analyses pre-specified during the design phase might not even take place because the trial is {\it stopped} early for either efficacy or futility. We discuss below what ``stopping a trial'' actually entails. 
%
%

We provide further considerations for the terms defined in Tables~\ref{tab: glossary_analyses1} and \ref{tab: glossary_analyses2} in the following subsections. Of note, the introduced terminology is endpoint specific. This will be discussed further in Section~\ref{sec:multipleEPs}. In Section~\ref{sec:general}, for the sake of simplicity, we only consider trials with a single primary endpoint.

\subsection{Primary analysis}
In the statistical literature the commonly used template term for what we call here {\it primary} analysis is {\it final} analysis, see e.g., Jennison and Turnbull~\cite{jennison_00} or Wassmer and Brannath~\cite{wassmer_16}. Hofner et al.~\cite{Hofner2023} already discussed the potential confusion among stakeholders when using the term {\it final} analysis. While motivated through construction of a group-sequential design it may generate the perception, especially among non-statisticians, that \textit{final analysis results} are \textit{final}, i.e., will not change anymore and that any \textit{interim results} are just \textit{preliminary} and subject to final confirmation. However, as long as we estimate a population quantity, results will never be \textit{final} (since we cannot collect infinitely many observations), meaning that in this context \textit{final} only refers to the last pre-planned analysis within the hypothesis testing framework at an information fraction of (about) 1. In trials with a time-to-event endpoint this is further aggravated by the fact that not all patients will have the event of interest during the trial. For example, only a fraction of patients will have a progression or die. Some patients might even be cured. To avoid confusion, Hofner et al.~\cite{Hofner2023} proposed to use {\it primary} analysis at the design stage and we reiterate this recommendation here.

\subsection{Confirmatory analysis}
Within a hypothesis test, the null hypothesis can only be rejected once for a given endpoint. Assume the null hypothesis for an endpoint is rejected at an {\it interim analysis} within a group-sequential design. We propose to refer to this analysis as {\it confirmatory} in the trial report and publications, instead of referring to it as an interim analysis. The term {\it interim analysis} is potentially misleading as it suggests incomplete information and that a subsequent {\it final analysis} might overwrite the hypothesis decision. If the hypothesis was rejected at the planned {\it primary analysis} we propose to call this the {\it confirmatory} analysis as well. If the hypothesis was not rejected at any of the pre-specified interim analyses nor the primary analysis, the primary analysis is the confirmatory analysis which, in effect, confirmed the lack of a statistically significant outcome for the trial. 
It is of paramount importance to plan the group-sequential design such that the information available at any interim analysis and hence at each potential confirmatory analysis is sufficient for the intended purposes, e.g., regulatory submission. In particular, sufficient safety data and sufficiently mature data on other important efficacy endpoints (e.g., OS if PFS was the primary endpoint) should be available to support the risk-dimension of a benefit-risk assessment.

\subsection{Updated analysis}\label{sec:updates}
Assume that the null hypothesis was rejected for an endpoint at an interim analysis or the primary analysis. {\it Updated analyses} at later \ccod{} do no longer have the purpose to \textit{test} that hypothesis for this endpoint. Rather their objective is {\it estimation}. The decision on the null at the confirmatory analysis persists: a null hypothesis can only be rejected once and significance can neither be lost nor enhanced at a later analysis. 

However, estimation of the underlying effect of interest can be improved at later \ccod{s} in the sense that: (1) the effect estimate will leverage more data as it includes more subjects and/or more events and/or longer follow-up meaning that (2) uncertainty around this estimate will therefore be reduced; and (3) the estimate may (better) capture time-dependent aspects of the treatment effect, such as a cure fraction or non-proportional hazards. However, the robustness of updated analyses always needs to be factored in, as there is potential for bias as a result of the confirmatory analysis and subsequent changes in trial conduct this may prompt -- intentionally or unintentionally. An example for an intentional change is a trial which allows for crossover to active treatment for all patients after the primary endpoint was met ("open label extension"). More subtle changes include modifications in the follow up or adjudication routines after the endpoint was met, or changes in the investigators' or patients' behaviour and beliefs after the endpoint was met.

While a null hypothesis has been rejected at a confirmatory analysis, an updated analyses could potentially impact the benefit-risk assessment, e.g., if updated data suggests a strong waning of (vaccine) efficacy or if detrimental effects in patient relevant endpoints are observed at a later point in time (e.g., positive PFS results paired with OS detriment). Updated results, on the other hand, can also support or strengthen the primary analysis in a meaningful way (e.g. numerical OS benefit after PFS was superior, or durable response to a treatment). As such, it should be of interest to all stakeholders to continue the trial unchanged for all planned updated analyses as long as possible, to reduce the risk of impeding interpretability of trial results. In addition, timing of analyses should be planned to ensure that sufficient data is available to address the questions of interest (i.e., survival benefit at a specific time-point instead of just a relative difference in the hazard functions). The importance of the "most informative" and mature data is also mentioned in multiple places in the (draft) guideline for the assessment of Section 5.1 of the Summary of Product Characteristics (SmPC)~\cite{ema_SmPC_5_1}, where it is requested that these data are presented to inform patients and prescribers.

For a binary or continuous endpoint measured at a fixed time point post randomization, the snapshot taken at the \ccod{} of the primary endpoint might actually already be {\it complete} in the sense that the endpoint is observed for every patient in the trial. This implies that updated analyses are not applicable for that endpoint at that given time point. 

Note that if the null hypothesis is rejected at an interim analysis, the analysis termed {\it primary} at the design stage loses its meaning. There is no \textit{obligation} to conduct such an analysis later when the pre-specified number of patients or events has been reached. Rather, the choice of \ccod{s} for updated analyses should be guided by {\it estimation} requirements, e.g., based on the desired precision of the estimate (width of the confidence interval), data maturity in terms of follow up time, or an important secondary endpoint. Further, if the null hypothesis is not rejected at an efficacy interim analysis, the interim analysis has no particular role related to the further analyses. These will occur as planned. 

There may be confusion between terms such as {\it updated}, {\it descriptive} and {\it exploratory} analyses. We propose to use the term {\it updated} for any analyses that are done after rejection of the null hypothesis within a group-sequential design. To the best of our knowledge there is no formal definition in the statistical literature for {\it descriptive} and {\it exploratory} analyses. We propose to reserve these terms for any analysis that is done outside of a formal hypothesis testing framework with type I error control and define an {\it exploratory analysis} as targeting hypothesis generation. It should still be pre-specified in the protocol though. We propose to use the term {\it descriptive analysis} for those analyses supporting the analysis of endpoints pertaining to formal hypothesis testing. 

\subsection{Decisive Analysis}

Typically, clinical trials should be planned such that the confirmatory analysis includes all the information that is needed by regulatory agencies, HTA bodies or other stakeholders for decision making. However, this does not always happen in practice either due to data driven aspects, e.g., more follow up is needed to characterize non-proportional hazards, or due to suboptimal planning of the analysis time points. In that case, updated analyses are an integral part of the assessment and for the interpretation of the derived results (see Section~\ref{sec:updates}). We propose to call the analysis that is used to make the benefit-risk assessment of an experimental treatment the {\it decisive analysis}. Typically, the decisive analysis coincides with the confirmatory analysis, but there might be exceptions where the decisive analysis happens later.

\begin{landscape}
\begin{table}[htbp]
\footnotesize
\centering
\begin{tabular}{|L{3cm}|L{8cm}|L{10cm}|} \hline
{\bf Term}                        & {\bf Design stage}   & {\bf Reporting stage} \\ \hline
Efficacy interim analysis     
&  In a trial with a group-sequential design one or more {\it pre-specified} interim analyses are performed under type 1 error control to determine whether the null hypothesis can be rejected. 
&   \multirow{2}{=}{Terminology no longer relevant at the reporting stage. Refer to an analysis as {\it confirmatory} or {\it updated} as applicable.} 
\\ \cline{1-2}
Primary analysis                 
& The last planned efficacy analysis in a group-sequential design under type 1 error control. Further analyses might be planned (see updated analyses) but are outside of type 1 error rate control.
& \\ \hline
Futility analysis                  
& A pre-specified futility analysis to assess whether the pre-specified stopping criteria for futility has been met.  
& If a trial is stopped for futility, refer to this analysis as {\it futility analysis}.\\ 
& \\ \hline
Confirmatory analysis                  
& Not applicable.  
& The analysis at which the null hypothesis is rejected or the last pre-specified analysis in case the null hypothesis was never rejected within the pre-specified hypothesis testing scheme. This can be an interim or the primary analysis at the design stage.\\ 
\hline
Updated analysis            
&  Any analysis after the confirmatory analysis. It is recommended to {\it pre-specify} such analyses (including their timing and purpose) at the design stage in the protocol. 
& Any analysis after the confirmatory analysis for a given endpoint that includes more data. This can include a pre-specified original interim or primary analysis (design stage), but also other analysis time points. It should be explicitly stated what analyses will be used as updated analyses. This may have to be defined conditionally on the time (interim analysis or primary analysis) when the null hypothesis for that endpoint was rejected, see Table~\ref{tab:flowchart}.  
\\ \hline
Decisive analysis  
&  Any confirmatory or updated analysis. It is recommended to {\it pre-specify} this analysis at the design stage in the protocol.
&  Analysis used by regulators, HTA bodies, and other stakeholders for their decision making (e.g., to make the benefit-risk assessment or cost-benefit assessment of a medicine). Typically this coincides with the {\it confirmatory} analysis, but there might be exceptions where the decisive analysis happens later.    
\\ \hline 
\end{tabular}
\caption{Glossary of terms for different analyses, Part 1. These definitions are usually endpoint specific.}
\label{tab: glossary_analyses1}
\end{table}
\end{landscape}

\begin{landscape}
\begin{table}[htbp]
\footnotesize
\centering
\begin{tabular}{|L{3cm}|L{8cm}|L{10cm}|} \hline
{\bf Term}                        & {\bf Design stage}   & {\bf Reporting stage} \\ \hline
Exploratory analysis               
& Truly exploratory in nature and could be hypothesis generating. Should be pre-specified in the protocol and can be based on data from a confirmatory or updated analysis.  & As for design stage. \\  \hline
Descriptive analysis               
& Descriptive analysis of endpoints to support hypothesis testing. For example, descriptive statistics and graphical presentation of data over time when hypothesis testing is focused on one time point. Should be pre-specified in the protocol and can be based on data from a confirmatory or updated analysis.   & As for design stage. \\  \hline
\end{tabular}
\caption{Glossary of terms for different analyses, Part 2. These definitions are usually endpoint specific.}
\label{tab: glossary_analyses2}
\end{table}
\end{landscape}

\section{Additional Clarifications on Group-Sequential Designs}
\label{sec:add}

The terminology above is relatively simple for a group-sequential design with a single hypothesis of interest. However, additional complexities arise in most situations in practice. These complexities may not be obvious in methodological developments which focus on single-endpoint situations, where all data is available as soon as a subject is randomized. In this section, we aim to clarify some additional subtle, yet still important, considerations to support development, discussion and implementation of well-designed group-sequential trials. 

\subsection{Multiple endpoints, arms, or subgroups}
\label{sec:multipleEPs}
Most confirmatory clinical trials aim for rejection of more than one null hypothesis. Frequently, there may be insufficient information available to do this for all endpoints at the same time. In such cases, it will be beneficial to define endpoint-specific sets of analysis time points,  e.g., \textit{interim} and \textit{primary analyses for PFS} and \textit{interim} and \textit{primary analyses for OS} in oncology trials. These definitions at the planning stage subsequently may lead to unique analysis time points for the respective \textit{confirmatory analyses} at the reporting stage, while the \textit{decisive analysis} is usually based on a single \ccod.

In a group-sequential design, a type 1 error control strategy will need to be considered for all endpoints with hypothesis testing \textit{individually} \cite{hung_07, glimm_10}---otherwise an inflation of the family-wise type 1 error may occur. We recommend to use endpoint specific terminology to facilitate this aspect.

Multiple hypotheses are also tested in designs with more than two arms with either confirmatory adaptive arm selection at an interim analysis (an example is the GATSBY trial, see Thuss et al.~\cite{thuss_17}), formal pairwise comparisons between all arms as in the CLL11 trial (see Asikanius et al.\cite{asikanius_16}), or potentially within a platform trial (see e.g. Howard et al.\cite{howard2018recommendations}). In such trials, comparisons may occur at different time points, thus requiring a different timing of analyses and hence also different \ccod{s} / \ssd{s} per comparison. In particular, an interim analysis for intervention A vs. C may become a confirmatory analysis, while evidence to claim superiority of B vs. C may only be available at a later interim (and then hence confirmatory) analysis. Hence, at reporting stage terminology such as {\it confirmatory analysis for A vs. B} and {\it confirmatory analysis for A vs. C} should be used.

Lastly, multiple hypotheses are tested in situations where multiple populations of interest are assessed, as in enrichment designs, see e.g., Jenkins et al.\cite{jenkins_11}. Here, analysis timings and \ccod{s} may be population specific. For example, timing of the primary analysis for population A may differ from the primary analysis for population B. 

 In all the above situations, the terminology introduced in Section~\ref{sec:general} can be easily generalized to become hypothesis specific.

\subsection{Pipeline information, overrunning and underrunning}\label{subsec:pipeline}

A key motivation for the development of this paper was that there is typically a gap between recruitment of study participants and observing their outcome. If all outcomes were available at the same time for all participants, there might be just one single analysis time point. Different terminology has been used in the past interchangeably to describe data which is accruing but has not been reported yet. This subsection aims to clarify the difference between pipeline information and (random) overrunning and how this possibly impacts trial design.

\subsubsection{Pipeline information}

{\it Pipeline information} is a term which is of particular relevance for trials with endpoints that are captured at a specific follow-up time after randomization. Examples of the former include the 6 minute walk distance at 26 weeks in Pulmonary Hypertension or the clinical remission rate at 52 weeks in Crohn’s Disease. 

Often, and especially at interim analyses, at the time of \ccod{} not all patients will have reached their targeted follow-up. For example, if an interim analysis is conducted when 60 of originally 100 planned patients have reached Week 26 (primary endpoint timing), while 80 have been enrolled, the data on the 20 patients who have been enrolled, but did not reach their primary endpoint yet, is referred to as pipeline information\cite{Hampson2013group}. This includes data between the \ccod{} and \ssd but also information beyond the \ssd. This data is usually not part of the analysis that is reported for this \ccod{}. The most common exception are pipeline deaths which are reported as part of the complete assessment of safety. However, it may be that pipeline data are deemed important later and used in the {\it decisive analysis} to update estimated effects received at the confirmatory analysis (be it an interim or the primary).

In contrast to this pipeline data, we define {\it under-} or {\it overrunning} to occur if an analysis is performed at an information fraction smaller or larger than what was planned at the design stage. Under- and overrunning result due to data being entered and cleaned between the \ccod{} and \ssd. In particular, this includes:
 \begin{itemize}
 \item events for time-to-event endpoints which have been entered after the \ccod{} but occurred prior,
 \item events resulting from data cleaning activities,
 \item removal of events which have not been confirmed in review, 
 \end{itemize}
 cf. Section~\ref{subsec:running_interim}. Under- and overrunning can also happen for other types of endpoints, whenever the amount of data exceeds or misses the originally targeted information fraction for the planned analysis. 

The EMA Reflection Paper on adaptive designs\cite{ema_adaptive} has a section on ``overrunning'', which is embracing {\it pipeline information} and {\it overrunning}, without spelling out a difference in interpretation and handling of the two. The Reflection Paper\cite{ema_adaptive} points towards decision making on {\it all available data}, thereby raising the methodological issue of potentially ``unrejecting'' a hypothesis or rejecting it twice. The reflection paper concludes the section with the point that decision making should be based on estimates of the treatment effect, instead of $p$-values alone. This concluding remark introduces ambiguity on regulatory acceptance of trial results and hence also uncertainty when designing and conducting group-sequential designs. The work by Hampson and Jennison\cite{Hampson2013group} is motivated by this ambiguity and offers an approach by defining the term {\it pipeline information} and handling the deterministic case of pipeline information within a pre-defined sequential testing framework. 

In general, trials (and duration of cleaning periods between \ccod{} and \ssd) should be planned in a way that the effect observed at any analysis is sufficiently mature and stable as not to relevantly change with additional (pipeline) observations.

\subsubsection{Over- and underrunning for a time-to-event endpoint at an interim analysis} \label{subsec:running_interim}

To appreciate why over- or underrunning the target number of events is a common challenge in trials with a time-to-event endpoint, one needs to understand how data is captured in the clinical trial database and how the \ccod{} is determined. 

The first observation is that data in large trials is always entered with a time-lag: for example, a progression event in an oncology trial is typically not entered on the day it was detected by the treating physician, but later when the center enters the data in the trial database en bloc. An additional source of delay is the data cleaning and potential adjudication. The sponsor or its data monitors checks the key data for plausibility, consistency and correctness, so that data might be subject to change, including changes to dates of events, or even the addition or removal of events. A third important factor is that in large multinational trials, prospective planning and communication of timelines is required because a large number of individuals are responsible for day-to-day trial conduct. A date for the \ccod{} is predicted based on the past occurrence of events. Because of variability (e.g. in how events happen) the number of observed events by that date will differ from the predicted, targeted number of events. 

At the \ssd, a snapshot of the cleaned database will be taken including all available data up to the \ccod. It typically happens that we do not precisely meet the target number of events. We either have fewer ({\it underrunning}) or more ({\it overrunning}) events than predicted. This results in an information fraction which is lower or higher than planned. Significance levels computed at the design stage based on the assumed information fractions are therefore recalculated according to the observed information fraction, where information fractions remain relative to the target number of events planned for the primary analysis. Recalculation is done using the $\alpha$-spending approach introduced by Lan and DeMets~\cite{lan_83}.

\subsubsection{Over- and underrunning for a time-to-event endpoint at the primary analysis} 
\label{subsec:running_primary}

As one can easily deduce based on the reasons for over- or underrunning at an interim analysis, this can also happen at the primary analysis. In that case, the information fractions are updated according to the actually observed maximum information. Updating the significance levels of all analyses based on these new information fractions using the pre-specified $\alpha$-spending function is not permissible, as it would require the modification to the critical boundaries for analyses which were already conduced at earlier interim analyses. The modification of these critical values and hence of those decisions would harm the validity of the trial. Instead, one uses the $\alpha$ that has actually been spent at earlier interim analyses and recomputes the remaining significance level for the primary analysis based on these and the updated information fractions. Wassmer and Brannath~\cite{wassmer_16} (Chapter 3.3) and Wolbers et al.~\cite{rpact_running} offer a comprehensive summary of this approach including further references. Obviously, this approach to under- and overruning is only valid as long as the under- or overrunning has occurred at random. Scenarios in which the team actively decides to delay the primary analysis due to missing success at an interim analysis would effectively result in a sample-size re-estimation. This approach would result in type I error inflation, unless proper adjustments are being implemented for correction of critical values or test statistics. 

Similar considerations on over- and underrunning apply for other types of endpoints, such as binary and continuous endpoints with fixed follow-up. However, the pipeline information available at the analysis time is (much) more predictable for those endpoints due to the known and fixed follow-up until read-out of the endpoint.

\subsection{Stopping a trial}\label{sec:stopping}

In discussing trial designs across stakeholders, the formulation {\it stopping for efficacy} and {\it stopping for futility} is often used. Indeed, if the null hypothesis corresponding to the primary endpoint can be rejected at an interim analysis for efficacy within a group-sequential design, then the testing for this hypothesis is completed. In statistical terms, the trial is ``stopped''. The result is {\it statistically significant} and the trial is formally a ``success". 

Similarly, if the trial is declared as futile, no further formal hypothesis testing will be conducted and the trial is considered ``negative'', even though it may have generated important insights. Notably, a futility declaration may not necessarily mean that the tested intervention is not efficacious, but could also point towards an inadequate trial design. 

From an operational perspective, the hypothesis test of the primary endpoint is not directly linked with how patients are treated and followed in a clinical trial. Indeed, data collection in the trial will typically continue after stopping for efficacy, i.e., rejection of the null hypothesis. In some cases, a trial simply continues to allow patients to still receive treatment and/or to collect more data to comprehensively describe and quantify the treatment effect, e.g., on secondary or safety endpoints\cite{day_07}. An example is in oncology, where PFS may be the primary and OS the secondary endpoint.

{\it Stopping a trial for efficacy} usually impacts the conduct of a trial even if the trial is not stopped from an operational perspective and further data is collected: unblinded summary measures and results might be released; the trial sponsor might be unblinded to individual patient data; investigators might be unblinded; patients might be unblinded and/or be allowed to cross-over to the new therapy. All this has the potential to introduce bias and impact the continued collection of reliable and relevant data. Hence, care should be taken at the planning stage to ensure that results at this stage a sufficiently mature for all important objectives. If collection of further information is essential, measures should be considered to mitigate bias if possible. These measures could include clear communication strategies, firewall procedures to protect the core trial team, and strategies how to continue the trial.

{\it Stopping for futility} has different implications and motivation. {\it No claim} is possible about whether a treatment effect is present or absent. It has simply been {\it decided} to stop the trial because the likelihood is too small that statistical significance will be reached at a later interim analysis or at the primary analysis. In contrast to an interim analysis for efficacy, if a trial is stopped for futility, operational activities in the trial are highly impacted. Patient recruitment stops, the trial is unblinded, treatment potentially discontinued, data collection is reduced to what remains necessary for safety of patients and regulatory surveillance, and the trial data is analyzed and results are published. Both, from a statistical and practical point of view, it is usually not possible to continue exposing new patients in the trial after the decision for ``futility" has been taken. 

A futility analysis reduces the power of a hypothesis test, and therefore also the probability of a type 1 error. Hence, one might be tempted to adjust the critical value of the hypothesis test such that the significance level is again fully exploited, implying a slight power increase. In that case it is imperative to stick to the futility decision as otherwise the type 1 error would no longer be controlled. As the decision to stop a trial for futility is usually multi-faceted and includes amongst others also a review of the safety data, non-binding futility analyses are generally preferable\cite{gallo_23}. As these authors explain, the term ``non-binding'' applies to both its semantic interpretation and its statistical definition: reaching such a threshold does not rigidly mandate stopping, and final efficacy thresholds are not loosened so that a decision to continue does not inflate the type I error. It is important to clearly keep the two meanings of {\it non-binding} apart.

\section{Discussion}
\label{sec:discussion}

Standardized terminology is important to ensure a common understanding of the objectives and scope of different analyses  in a clinical trial among all stakeholders. In this paper, as a team of statisticians from pharmaceutical industry, regulatory agencies, and academia, we propose terminology for clinical trials with interim analyses and confirmatory intent, i.e. hypothesis testing. The nomenclature can, however, be used flexibly in different types of clinical trial designs, including group sequential trials with multiple endpoints. 

Firstly, it is important to be clear on the scope of the interim analysis, whether it is futility or efficacy or both and what exactly is meant when trialists say ``a trial is stopped'' as a result of an interim analysis. We distinguish between the {\it design} and {\it reporting} stage because the purpose and terminology of analyses at the reporting stage depends on the results which are, naturally, not known at the design stage. Importantly, we think that if the null hypothesis was rejected at an interim analysis, it should not be called an interim analysis thereafter because it reflects the confirmatory analysis of the trial. Furthermore, from an inferential point of view, there is no difference whether the hypothesis is rejected at an interim analysis or primary analysis at the initially targeted maximal number of patients or events. Hence, there should be no difference in nomenclature. We also propose that the analysis time point based on the target number of patients or events should not be called the final analysis, as commonly referred to, because if the trial was positive at an interim analysis the null hypothesis will not be tested again for the same endpoint and, in a trial with a time-to-event endpoint, the patients are followed past this analysis time point. Hence, it is misleading to call such an analysis ``final''. {\it Updated} analyses after the confirmatory analysis might furthermore be performed after the primary analysis, with the primary purpose of reducing uncertainty in estimation of relevant quantities. However, the reduction in uncertainty has to be weighed against potential introduction of bias through unblinding or changes in the treatment after the confirmatory analysis. For example, it might be allowed that patients cross over to the active treatment after it was proven effective.

To design trials well, factors other than type 1 error control also need to be considered, such as sufficient follow-up and the potential concrete actions taken following each analysis. These should be clearly and transparently outlined in the protocol. At the reporting stage, critical values need to be adjusted based on the observed number of events.

The concepts discussed and explained in this paper are generalizable to many trial designs and type of endpoints. There are specific situations where implementation may not be as straightforward as described here. For example in case of non-proportional hazards the treatment effect is not constant over time, e.g., in an indication where part of the patients are cured by the treatment. This needs to be considered when planning feasibility of, especially futility interim analyses. The terminology is, however, equally relevant.

To conclude, standardized terminology facilitates clear and transparent communication which is essential during planning, conduct and reporting of clinical trials. Increasing complexity and variety in clinical trial designs further enforces the need for standardized terminology, careful planning, and clear communication among all relevant stakeholders.

\section{Declarations}

\subsection*{Ethics approval and consent to participate}

Not applicable.

\subsection*{Consent for publication}

Not applicable.

\subsection*{Availability of data and material}

Code to reproduce all the computations in this paper is available as a quarto file on GitHub\cite{asikanius_24_code}.

\subsection*{Competing interests}

Not applicable.

\subsection*{Funding}

Not applicable.

\subsection*{Authors' contributions}

EA, BH, TM, CUK, KR conceived the idea for article and drafted it. All authors critically reviewed the manuscript and approved its final version.

\subsection*{Acknowledgements}

Not applicable.

\subsection*{Authors' information (optional)}

Not applicable.

\bibliographystyle{abbrv}

\bibliography{ref}

\appendix

\section{Description of a group-sequential design in a trial protocol or statistical analysis plan}
\label{sec:example_protocol}

To conclude this paper we provide illustrative text that could be included in a trial protocol or statistical analysis plan. Note that this is again a hypothetical example only and choices might differ in practice. The reporting of that trial will be discussed in Section~\ref{sec:trialresults} below. Note that, other than in this example, in a real trial it would be expected to justify the assumptions and design choices.
As this section, as well as Section~\ref{sec:trialresults}, may serve as templates we do not use any abbreviations on purpose.

\begin{quote}

\bigskip

    {\bf Determination of number of events}
    
    The primary endpoint of this trial is overall survival. Randomization will be 1:1 between experimental arm and control. An overall one-sided significance level of 0.025 and a power of 80\% power will be assumed. Interim analyses for 
    \begin{itemize}
    \item futility will be performed after 1/3 and 2/3 of events, 
    \item efficacy will be performed after 2/3 of events.
    \end{itemize}
    The type I error will be controlled adjusting for the two planned analyses for efficacy using the Lan-DeMets approximation to the O'Brien-Fleming boundary shape. Under these assumptions, 385 events are required to have 80\% power to reject the null hypothesis that hazard ratio $\geq 1$ when the true hazard ratio = 0.75. Information fractions, the targeted number of events and the predicted time of reaching these events (analysis time) are provided together with nominal significance levels, and critical values (on the hazard ratio scale) for statistical significance in Table~\ref{tab:trial1}.
    Futility boundaries are chosen based on clinical grounds. Overall power loss for adding the two futility interim analyses is about 2\% and will not be accounted for in the computation of the targeted number of events.

\begin{table}[htbp]
\centering

\begin{tabular}{|c|c|c|c|c|c|c|}\hline
\multirow{2}{1.1cm}{Event} & \multirow{2}{2.4cm}{Information\linebreak fraction} & \multirow{2}{1.3cm}{\hspace*{0.1cm}Events}  & \multirow{2}{2.5cm}{Analysis time (months)} & Futility bound & \multicolumn{2}{c|}{Efficacy bound} \\ \cline{5-7}
          & & && Hazard ratio & $\alpha$ & Hazard ratio \\ \hline
First patient in & 0 & 0 & 0  &  &  &  \\ \hline
Interim analysis 1          & 0.33 & 129 & 19.7  & 1 & &    \\ \hline
Interim analysis 2          & 0.67 & 257 & 35.6  & 0.9 & 0.012 & 0.731  \\ \hline
Primary analysis & 1.00 & 385 & 54.8  &                            & 0.046 & 0.816  \\ \hline 
Updated analysis     &  & 500 & 76.4  &  &  &  \\ \hline
\end{tabular}

\caption{Design details of the hypothetical trial. The significance level for the efficacy bound is 2-sided. Empty cells denote that this value is not applicable.}
\label{tab:trial1}
\end{table}
    
\bigskip

    {\bf Timing of analyses}

    Analysis timings are computed based on the following assumptions:
    \begin{itemize}
        \item 100 patients will be recruited per month over a period of 12 months, giving a total number of 1200 patients. 
        \item We assume 2.5\% yearly dropout in both arms, with an exponential time-to-dropout distribution.
        \item We assume exponentially distributed survival times for the primary endpoint, with a median overall survival of 6 and 8 years on the control and experimental treatments, respectively (which implies a hazard ratio of $6 / 8 = 0.75$).
    \end{itemize}
    With these assumptions it is predicted to take 19.7, 35.6, 54.8, and 76.4 months from first-patient-in to the clinical cutoff date of the two interim, the primary, and the updated analysis, respectively.
    
\bigskip

    {\bf Expected follow-up}

The last patient will be randomized after 12 months, so minimal follow-up will be approximately 7.7, 23.6, 42.8, 64.4 months at the four pre-specified analyses given in Table~\ref{tab:trial1}, with up to additional 12 months follow-up.

\bigskip

    {\bf Update significance levels with observed information fractions}

    Using the Lan-DeMets approximation to the O'Brien-Fleming boundary shape, local significance levels will be re-calculated based on the observed information fraction at each analysis.

\bigskip

     {\bf Designation of analyses}
   
    At the design stage, names of analysis are given as in the first column of Table~\ref{tab:flowchart}. The label of the analysis at the analysis stage depends on the outcome of the trial (see Table~\ref{tab:flowchart}). If the trial stops at interim analysis 1 or interim analysis 2 for futility, the corresponding analysis is also referred to as futility analysis at the reporting stage. If the trial 
    \begin{itemize}
        \item runs to interim analysis 2 and stops there for efficacy, or
        \item runs to the primary analysis 
    \end{itemize}
    this last analysis will be labelled as the {\it confirmatory analysis} for reporting purposes, regardless of its outcome. 
    
    In the case of futility, an analysis at the time of the primary analysis will be conducted as \textit{updated analysis} to obtain the relevant safety information for reporting purposes and subsequent trial planning. No further \textit{update} will be conducted in that case. 
    
    In case of early stopping for efficacy, the primary analysis will no longer be conducted. If the trial was significant in any of the analyses, an \textit{updated analysis} will be conduced after 500 events have been reached or after a minimum follow up of six years was reached for each patient in order to get a more precise estimate and a longer follow up. The time point of six years was chosen to be later than the 5.3 years at which 500 events were initially predicted. This choice was made to collect as many events as possible to but to restrict the duration of the trial to a reasonable time-frame in case of slower event accrual than anticipated. The primary analysis will not be executed if the trial was significant at interim analysis 2. 

\bigskip

    {\bf Blinding}

    The trial is triple-blind, i.e. neither patients nor treating physicians nor the sponsor's trial team know the administered treatment. An independent Data Monitoring Committee is set up to conduct periodic safety reviews (see e.g. \cite{schoeffski_21} for details on setting up an iDMC). The iDMC is also responsible for evaluating efficacy data at the two pre-specified interim analyses in order to protect the trial integrity and blinding with respect to treatment allocation and treatment-group specific effects. The data for the iDMC meetings are prepared by an independent, external statistician. A firewall is in place to separate unblinded information from the sponsor's trial team. 
    
    The trial team remains blinded to the treatment allocation of patients until futility or efficacy is declared at an interim analysis or until the primary analysis. Patients and investigators remain blinded to the individual treatment assignment until the end of the trial, unless futility was declared, or until disease progression.
    
    In case, at a futility analysis the experimental treatment is considered to be futile, enrolment is stopped immediately. Patients already enrolled in the trial are unblinded. Patients on control treatment will be continued on their treatment as per protocol. Already enrolled patients on experimental treatment may continue on their treatment upon investigator's and patient's discretion unless it is specifically recommended to discontinue treatment immediately. Collection of overall survival data will be continued until the updated analysis. All other data collection activities will be reduced to the minimum, e.g. safety data will be collected as required\footnote{The protocol should clearly outline the potential outcomes of each analysis both at study level and at individual patient level. For example, the protocol should outline whether ongoing treatment is immediately stopped for all patients in case of a futility stop or whether only further enrollment is discontinued and how data collection is impacted. Here we give one possible example.}. Further details and responsibilities of the iDMC and the trial team are defined in the iDMC Charter.

\begin{landscape}

\begin{table}[htbp]
\centering
\begin{tabular}{|C{4cm}|C{3.5cm}|C{3.5cm}|C{3.5cm}|C{3.5cm}|} \hline
\multirow{2}{4cm}{\centering \bf Design stage} & \multicolumn{4}{c|}{\bf Reporting stage} \\
\cline{2-5}
                                        & {\bf Stop at IA1}           & {\bf Stop at IA2 for futility} & {\bf Stop at IA2 for efficacy} & {\bf Stop at\newline primary analysis} \\ \hline
\bf IA1\newline  (futility)             & {\it Futility analysis} &  & & \\ \hline
\bf IA2\newline (futility \& efficacy)  &                             & {\it Futility analysis} & {\it Confirmatory analysis} & \\ \hline
\bf Primary analysis                    & {\it Updated analysis}      & {\it Updated analysis} & & {\it Confirmatory analysis} \\ \hline
\bf Updated analysis\newline & & & \multicolumn{2}{c|}{\it Updated analysis}  \\ \hline
\end{tabular}
\caption{Potential outcomes and names of corresponding analyses for the trial in Table~\ref{tab:trial1}. The {\it decisive analysis} can either coincide with the {\it confirmatory} or an {\it updated} analysis.}
\label{tab:flowchart}
\end{table}

\end{landscape}
\end{quote}

\section{Description of the results of a group-sequential design}
\label{sec:trialresults}

In this section we propose how to report the outcome of a trial that was stopped for futility or efficacy, respectively. To this end, we have simulated trials based on the assumptions of Section~\ref{sec:example_protocol} and under two scenarios for the treatment effect: Trial 1 with an assumed hazard ratio of 1 and Trial 2 with an assumed hazard ratio of 0.75. We repeatedly simulated new trials until we had a trial that was stopped for fultility at interim analysis 1 (Trial 1) and one which was stopped for efficacy at interim analysis 2 (Trial 2). These simulated trials are used as the basis for suggested reporting in Sections~\ref{subsec:reporting_futility} and \ref{subsec:reporting_efficacy}. 

When reporting results of a group-sequential design, or more general adaptive design, we strongly recommend to also follow the CONSORT extension for adaptive designs\cite{dimairo_20}, as applicable.

\subsection{Reporting of a trial that was stopped for futility}\label{subsec:reporting_futility}

For Trial 1, an interim analysis for futility only was pre-specified in the protocol to take place after 129 events with a boundary on the hazard ratio scale of 1 (see Table~\ref{tab:trial1}). The clinical cutoff date of this analysis was 01 November 2021 (i.e., 18.3 months after first patient in on 23 April 2020). The snapshot was taken four weeks later on 03 December 2021 and the independent data monitoring committee meeting took place on 18 December 2021. A hazard ratio of 1.07 based on 132 events was observed at analysis, which became confirmatory. Based on the presented data, the independent data monitoring committee decided to recommend to stop the trial for futility and this recommendation was endorsed by the sponsor on 19 December 2021.

\subsection{Reporting of a trial that was stopped for efficacy}\label{subsec:reporting_efficacy}

For Trial 2, an interim analysis for efficacy was pre-specified in the protocol to take place once 257 events had been observed with a threshold for efficacy (i.e., a minimal detectable effect) of hazard ratio = 0.731, corresponding to a nominal significance level of 0.012 (see Table~\ref{tab:trial1}). The clinical cutoff date of this analysis was 10 April 2023 (i.e. 34.7 months after first patient in on 23 April 2020). The snapshot was taken seven\footnote{Usually, cleaning for an efficacy interim analysis takes longer compared to an interim for futility, because if we stop at an interim analysis for efficacy the data needs to be fit-for-purpose to be filed. That is not the quality needed to make a futility decision.} weeks later on 30 May 2023. The independent data monitoring committee meeting took place on 05 June 2023.  

At the clinical cutoff date, 255 events had been observed, corresponding to an information fraction of 0.662. Using the Lan-DeMets approximation to the O'Brien-Fleming $\alpha$-spending boundary, the nominal significance level was recalculated to be 0.0117, thus updating the critical hazard ratio threshold from 0.731 to 0.729. 

The observed hazard ratio at this analysis, which became confirmatory, was based on 255 events and was 0.689 with unadjusted 95\% confidence interval from 0.535 to 0.881. Based on the presented data, the independent data monitoring committee recommended to stop the trial for efficacy. This was endorsed by the sponsor on 22 May 2023. Patients and investigators remain blinded to the to the individual treatment assignment and the trial continues for an updated analysis after 500 events or 6 years of minimum follow up have been reached.

It is known that theoretically, naive effect estimates from a trial that has stopped early are biased because of the selective nature of the sampling procedure\cite{wassmer_16}. Except in extreme situations this bias is not practically relevant\cite{freidlin_09, wang_16}. Indeed, computing a median unbiased estimate taking the group-sequential design into account\cite{wassmer_16} yielded an effect estimate of 0.691 with adjusted 95\% confidence interval from 0.540 to 0.883.

\end{document}